\documentclass[sigconf]{acmart}

\AtBeginDocument{%
  \providecommand\BibTeX{{%
    \normalfont B\kern-0.5em{\scshape i\kern-0.25em b}\kern-0.8em\TeX}}}

\acmYear{2024}\copyrightyear{2024}
\setcopyright{rightsretained}
\acmConference[ACM FAccT '24]{ACM Conference on Fairness, Accountability, and Transparency}{June 3--6, 2024}{Rio de Janeiro, Brazil}
\acmBooktitle{ACM Conference on Fairness, Accountability, and Transparency (ACM FAccT '24), June 3--6, 2024, Rio de Janeiro, Brazil}
\acmDOI{10.1145/3630106.3658910}
\acmISBN{979-8-4007-0450-5/24/06}
\newcommand{\ie}{\textit{i.e.,}}
\newcommand{\eg}{\textit{e.g.,}}

\usepackage{tabularx}

\usepackage{pifont}

\usepackage{dblfloatfix}
\usepackage{enumitem}
\setlist[itemize]{leftmargin=*}

\newcommand{\caseno}{40}
\newcommand{\abandonedno}{24}
\newcommand{\globalno}{16}
\newcommand{\owner}{algorithm owner}
\newcommand{\Owner}{Algorithm owner}

\usepackage{scalerel}

\usepackage{soul}

\begin{document}

\title[The Fall of an Algorithm]{The Fall of an Algorithm:\\ Characterizing the Dynamics Toward Abandonment}

\author{Nari Johnson}
\email{narij@andrew.cmu.edu}
\affiliation{%
  \institution{Carnegie Mellon University}
  \country{USA}
}

\author{Sanika Moharana}
\email{smoharan@andrew.cmu.edu}
\affiliation{%
  \institution{Carnegie Mellon University}
  \country{USA}
}

\author{Christina N. Harrington}
\email{charring@andrew.cmu.edu}
\affiliation{%
  \institution{Carnegie Mellon University}
  \country{USA}
}

\author{Nazanin Andalibi}
\email{andalibi@umich.edu}
\affiliation{%
  \institution{University of Michigan}
  \country{USA}
}

\author{Hoda Heidari}
\authornote{Both senior authors contributed equally to this research.}
\email{hheidari@cmu.edu}
\affiliation{%
  \institution{Carnegie Mellon University}
  \country{USA}
}

\author{Motahhare Eslami}
\authornotemark[1]
\email{meslami@andrew.cmu.edu}
\affiliation{%
  \institution{Carnegie Mellon University}
  \country{USA}
}

\renewcommand{\shortauthors}{Johnson, et al.}

\newcommand{\xhdr}[1]{\vspace{1mm} \noindent\textbf{#1.}}
\newcommand\extrafootertext[1]{%
    \bgroup
    \renewcommand\thefootnote{\fnsymbol{footnote}}%
    \renewcommand\thempfootnote{\fnsymbol{mpfootnote}}%
    \footnotetext[0]{#1}%
    \egroup
}
\begin{abstract}
As more algorithmic systems have come under scrutiny for their potential to inflict societal harms, an increasing number of organizations that hold power over harmful algorithms have chosen, or were required under the law, to \emph{abandon} them. While social movements and calls to abandon harmful algorithms have emerged across application domains, little academic attention has been paid to studying abandonment as a means to mitigate algorithmic harms. In this paper, we take a first step towards conceptualizing ``\emph{algorithm abandonment}'' as an organization’s decision to stop designing, developing, or using an algorithmic system due to its (potential) harms. We conduct a thematic analysis of real-world cases of algorithm abandonment to characterize the dynamics leading to this outcome. Our analysis of \caseno{} cases reveals that campaigns to abandon an algorithm follow a common process of six iterative phases: \emph{discovery, diagnosis, dissemination, dialogue, decision,} and \emph{death}, which we term the \emph{6 D's of abandonment}. In addition, we highlight key factors that facilitate (or prohibit) abandonment, which include characteristics of both the technical and social systems that the algorithm is embedded within. We discuss implications for several stakeholders, including proprietors and technologists who have the power to influence an algorithm’s (dis)continued use, FAccT researchers, and policymakers.
\end{abstract}





\maketitle

\section{Introduction}
An increasing number of incidents have demonstrated how algorithmic systems can inflict harm, by (for example) upholding oppressive power structures \citep{noble2018algorithms, benjamin2019race}, reinforcing stereotypes \citep{thevergeTwitterTaught, andalibi2023conceptualizing}, or making life-altering decisions through inscrutable logic \citep{selbst2018intuitive,deliban2018comment,forrest2021when,eppink2023testimony}.
In response to these harms, a growing body of research has explored how harmful algorithms might be \emph{repaired} through technical or procedural interventions \citep{schwennessen2019algorithmic,bembeneck2021playbook,orphanou2022mitigating}. 
However, scholars and activists alike have argued that not all algorithmic harms can be effectively addressed or mitigated through these kinds of repairs \citep{harcourt2006against,milner2020abolish,wang2023against}.
This stance often arises from a recognition of the deeply ingrained nature of the corresponding harm, where mere adjustments fail to resolve fundamentally systemic issues.
In some instances, resistance from impacted communities has precipitated the \emph{fall} of an algorithmic system, in which its proprietors might choose, or be forced to, \emph{abandon} its use. 

Take, for example, a case from Arkansas: when a new resource allocation algorithm resulted in significant cuts to the number of in-home care hours that Medicaid beneficiaries would receive \citep{btah2016arkansas}, one beneficiary wrote in with a clear request: to turn it off, and return to the old system of human-based assessment. 
No one in the state could offer an explanation for why this beneficiary's in-home care hours were reduced, cuts that for many beneficiaries with chronic disabilities resulted in drastic changes to their quality of life. 
While three years later, the algorithm (called RUGS) was abandoned, the road to this outcome was long and tedious. 
Efforts to abandon the algorithm were met with resistance from state officials \citep{weil2023using}. 
The eventual decommissioning of RUGS after it was invalidated by a state court \citep{btah2016arkansas} was a result of concerted advocacy efforts by impacted individuals, legal advocates, and media attention.

\begin{figure*}[t]
\centering
\includegraphics[width=1\linewidth]{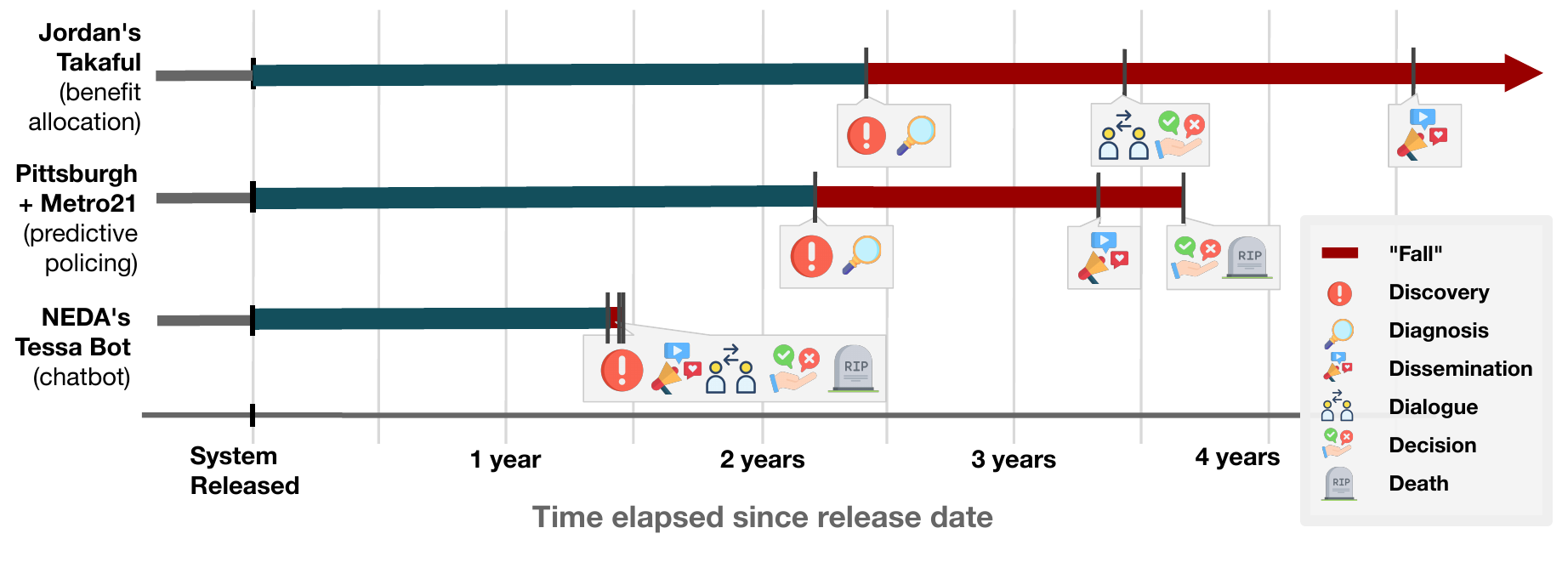}
\caption{Event timelines for three example algorithms \citep{hrw2023automated,capp2020primer,hoover2023eating}. We visualize the time elapsed to each event since each algorithm's release date (x-axis), and label important events by tagging them with one (or multiple) phase(s) from the 6 D's of abandonment (see legend, bottom right). We provide detailed descriptions of each event in Appendix \ref{apdx:timeline-descriptions}. Two of the pictured algorithms (the city of Pittsburgh's predictive policing pilot and the National Eating Disorder Association's ``Tessa'' chatbot) were abandoned, and had a ``Death'' phase. Takaful, a cash transfer allocation algorithm deployed in Jordan, was \emph{not} abandoned and is still in use at the time of writing (denoted by the red arrow into the present). These timelines illustrate how the time elapsed between events (\eg{} for the algorithm to be discovered, or be abandoned after discovery) varied considerably between the three cases shown.}
\label{fig:timelines}
\end{figure*}

Although some campaigns to abandon harmful algorithms have been successful, other harmful algorithms continue to remain in use today.
Furthermore, while some algorithms were abandoned in early ideation phases \emph{before} they were deployed, and some were abandoned within mere hours of harm discovery, other algorithms continue(d) to inflict harm for years.
Thus, open questions remain when considering the possibility of abandoning harmful algorithms as a mechanism of harm mitigation and accountability:
What types of events preceded and precipitated the decision to abandon an algorithm?
How did impacted communities and other actors work together to contest harmful algorithms?
What technical, political, or social characteristics of real-world algorithmic deployments posed barriers for contestation, accountability, and abandonment?
How can future calls for abandonment be bolstered and supported?

In this paper, we take a first step toward answering these questions by exploring the concept of \emph{algorithm abandonment}.
We define algorithm abandonment as a decision made by actors with jurisdiction over the system, to discontinue the process of developing, deploying, or using the algorithm due to its (potential) harms.
The decision to abandon an algorithm can be made at any point in the algorithm's development lifecycle \citep{btahlifecycle,suresh2021framework}, and includes both decisions not to build a proposed technology, and decisions to stop using tools that are already built.
The algorithmic systems that we explore span diverse ways of coming into being, including hand-coded rules and machine learning models. 
Our study of algorithm abandonment draws from literature from Science and Technology Studies which examines algorithms not just as technical artifacts mapping inputs to outputs, but rather as socio-technical ``assemblages'' of social relationships that enable and sustain technologies \citep{schwennessen2019algorithmic}. 
Thus, understanding abandonment requires examining the social, political, and economic systems that algorithms are embedded within \citep{seaver2019field}.

In our thematic analysis of \caseno{} different campaigns to abandon harmful algorithms in recent years, we observe that the fall of an algorithm (or ``\emph{dynamics}'' leading to abandonment) often follows a consistent pattern of six iterative, non-linear phases which we term the 6 ``D''s of abandonment: (1) \emph{discovery}, (2) \emph{diagnosis}, (3) \emph{dissemination}, (4) \emph{dialogue}, (5) \emph{decision,} and (6) \emph{death}, which we visualize for three example cases in Figure \ref{fig:timelines}.
We review each of these  phases 
and highlight both tactics that \owner{}s used to avoid scrutiny (\eg{} discrediting or ignoring criticism), and tactics that other actors such as impacted individuals, advocacy organizations, journalists, or regulators used to shift power to advocates of abandonment.
We additionally identify key \emph{socio-technical factors} 
relevant to the outcome of whether an algorithm was ultimately abandoned (and how long it took to reach this stage), such as public awareness of and access to the algorithm, 
characteristics of the algorithm's implementation, 
visibility of criticism, 
the regulatory environment, 
and for whom (and how) the algorithm is marketed.
We conclude by discussing implications for technologists designing algorithmic systems, regulators and oversight entities, and the FAccT community in identifying barriers to abandonment and interventions to support future calls to abandon harmful algorithms.
We release our annotated case set online\footnote{https://njohnson99.github.io/fall-of-algorithm-database/} as a resource to aid future research and advocacy efforts.

\vspace{-2mm}
\section{Background \& Related Work}

We contextualize our study of abandonment using scholarship that studies how different stakeholders respond to algorithmic harm. ``Algorithmic harm,'' as described by \citet{shelby2023socio}, refers to ``\emph{the adverse lived experiences resulting from a system's deployment and operation in the world}'' which occurs through the interplay of technical components and societal power dynamics  
\citep{noble2018algorithms, benjamin2019race, mcgregor2020preventing}.
We group approaches taken to address algorithmic harm into two broad categories: (1) actions taken ``from above'' by organizations in positions of power (\eg{} those that develop, oversee, or govern the algorithm's use), and (2) actions taken ``from below" by other stakeholders (\eg{} those impacted by algorithmic harm). 
We adopt the language of ``above'' and ``below'' from \citet{irani2021hci}, who uses these terms to articulate the ``\emph{tensions between the wants, needs, and knowledge of unequally empowered stakeholders}''.

We situate our study of abandonment as an organizational decision made ``from above'' by \owner{}s or government oversight entities. We use the term ``\owner{}'' to broadly refer to the organizations and people within them that have the power to influence an algorithm's (dis)continued use on the ground. 
Our analysis of the events, or ``dynamics'', that precede abandonment in practice draws from the growing body of scholarship that examines how actors ``from below'' organize to contest harmful algorithmic systems \citep{zong2023data, devrio2023building}. We use the term ``impacted communities'' to describe the people who are affected by the use of an algorithm. Those impacted may include the direct consumers of a technology or any other people who are subject to an algorithm's decisions \citep{friedman2006value, moss2021assembling}. 

\xhdr{Repair from above} Much work concerning algorithmic harm explores the actions that \owner{}s might take to repair flawed technologies. 
The AI/ML and FAccT community have proposed numerous technical methods to promote values of ``fairness'' and ``explainability'' \citep{doshivelez2017rigorous, selbst2018intuitive, liao2023ai}. This work has led to the development of metrics \citep{speicher2018unified, pessach2022review, barocas2023fairness, subhash2022what}, methods \citep{dwork2012fairness,wang2015falling, hardt2016equality, ribeiro2016why}, and toolkits \citep{bellamy2018ai, arya2019explanation, deng2022exploring, weerts2023fairlearn} that practitioners can use to measure and fix specific harmful behaviors. 
Beyond technical repairs, human-computer interaction researchers have documented how organizational users and decision-makers engage in their own ad-hoc or improvisational repair activities when they encounter flawed technologies \citep{houston2016values, eslami2017be, schwennessen2019algorithmic}. For example, social workers using AI tools to screen calls can override them to mitigate disparities in final outcomes \citep{deartega2020case, cheng2020heterogeneity, cheng2022how}. 
Such methods can be used to correct the algorithm's flaws, often defined in terms of narrowly scoped measures \citep{jacobs2021measurement}, while maintaining the algorithm's operation.

In both research and policy proposals, repair has emerged as a dominant approach to address algorithmic harm from above \citep{lee2019algorithmic, davis2021algorithmic, bembeneck2021playbook}. 
Both governments and private-sector organizations have instituted policies requiring mitigation strategies to repair potential algorithmic harms \citep{canadianaia, microsoft2022rai}. 
However, not all algorithmic harms are readily fixable, particularly in cases where critics take issue with the very existence of the system itself. 
Critical scholarship has scrutinized the use of algorithms in specific domains (\eg{} predictive policing \citep{khan2021police, milner2020abolish} or criminal sentencing \citep{forrest2021when}), and questioned if such algorithms can ever be repaired. 
Instead of examining how flawed technologies might be repaired, we focus our inquiry on abandonment, which is predicated on questioning if an algorithm should exist in the first place. 


\xhdr{Refusal from below} The concept of refusal, defined simply as the act of saying ``no'' \citep{ahmed2017no}, has been studied in feminist and de-colonial discourses \citep{simpson2007ethnographic, tallbear2013native, mcgranahan2016theorizing, honig2021feminist}. Today, scholars and activists continue to draw from these traditions in studying refusal of emerging technologies \citep{cifor2019feminist, devrio2023building, zong2023data}.
Among theories of anti-techno-solutionism, \citet{benjamin2016informed} refers to indigenous refusal to analyze vignettes where research subjects refuse to engage with new biotechnologies. 
A group of scholars drafted the Feminist Data ``Manifest-No'', a declaration of refusal to harmful data regimes and exploitative research practices \citep{cifor2019feminist}. 
There is a rich history of grassroots social movements formed to refuse technologies, tracing back hundreds of years to the British Luddites during the Industrial Revolution \citep{conniff2011what}, sparking modern adaptations of ``Luddism'' by workers organizing to oppose harmful algorithmic management practices \citep{chayka2023rethinking, merchant2023blood}.

Refusal of harmful technologies may take the form of individual users choosing to no longer interact with harmful algorithms \citep{devrio2023building, zong2023data}. For example, in response to Amazon's introduction of AI-powered cameras for worker surveillance, some drivers engaged in an act of refusal by obscuring the cameras with stickers, rendering them unable to record \citep{gurley2021amazon}. But refusal also extends beyond individual acts into collective calls for a system to be abandoned. Understanding calls to abandon an algorithm through the lens of refusal sheds light on the possibilities that abandonment entails to ``\emph{open up possibilities for power shifts}'' \citep{casemajor2015non}. 


\xhdr{Conceptualizing \& contextualizing abandonment} In this paper, we conceptualize algorithm abandonment as ``a decision made \emph{from above} (\eg{} by algorithm proprietors, vendors, purchasers, or governance bodies), to discontinue developing, deploying, or using a harmful algorithmic system, usually due to calls \emph{from below}". 
We focus our study on abandonment as an \emph{organizational} decision (rather than that of individual user decisions) as many systems are implemented in a way where opting out is difficult or impossible \citep{lecher2018what, tacids2021automated, amnesty2022ban}.
We differentiate abandonment from repair in that while repair considers how one might improve an existing algorithmic system, abandonment entails doing away with specific types of automation altogether. This conceptualization of abandonment builds upon scholarship \citep{baumer2011when, dombrowski2016social, homewood2019inaction, howell2021cracks, stapleton2022imagining} focused on ``\emph{when the implication is not to design ([a] technology)}'' and advocates for technologists to adopt ``\emph{a reflective awareness for when [automation] is inappropriate or harmful}'' \citep{baumer2011when}. 
\vspace{-5pt}


\section{Methodology}

We adopt a case study approach to analyze documented cases where stakeholders called for abandonment in recent years. 
Our goal was to understand the dynamics of algorithm abandonment by studying the events that occurred throughout these calls. 
As such, we first gathered real-world cases of algorithms that were (called to be) abandoned, and then conducted an iterative thematic analysis.

\xhdr{Case collection}
To develop our dataset, we conducted an extensive search of various sources for instances where a harmful algorithm was either (a) already abandoned, or (b) critics \emph{called for} the algorithm to be abandoned, regardless of whether it was abandoned or not at the time of writing. Our search methodology was a dynamic, iterative process that unfolded concurrently with our analysis. This approach involved consulting a diverse group of resources, detailed further below:
\begin{itemize}
    \item 
\emph{Initial dataset}: We began with an initial set of cases that the authors were familiar with from research articles and reading lists focused on resisting and refusing algorithmic harm  \citep{refusalconf, shen2021everyday, shelby2023socio, zong2023data, devrio2023building}. 
We also conducted web searches with keywords such as ``algorithm abandonment'', ``algorithmic bias'', and ``algorithmic harm''.

\item \emph{AI incident databases}:
We reviewed all of the reported incidents to date\footnote{We conducted our review from July to December 2023, and reviewed all incidents that had been posted in each database before January 1, 2024.} in three incident databases that compiled and documented real-world harms caused by algorithmic systems:
(a) the AI Incident Database (AIID) 
\citep{mcgregor2020preventing}, (b) the AI, Algorithmic, and Automation Incidents and Controversies (AIAAIC) 
repository \citep{aiaaic}, and (c) Not My AI 
\citep{varon2022not} which focuses specifically on mapping cases of algorithmic harm in Latin America.

\item 
\emph{Expert consultations}: We also identified additional cases through conversations with experts who had academic or lived experiences relating to algorithm abandonment.
Specifically, our research team met with members of an independent U.S.-based advocacy group founded to help impacted communities resist oppressive surveillance technologies.
We also met with academic colleagues who have done research on topics relating to resisting and responding to algorithmic harm.
These experts pointed us to relevant cases that they had encountered in their work.

\item 
\emph{International cases}: Given that U.S. and European-based incidents were over-represented in our initial case set, we extended our search to be intentionally inclusive of other regions globally.
We conducted additional keyword searches on Google to identify relevant cases that occurred in countries outside of Western contexts, such as Korea \citep{kwon2021ai}, Argentina \cite{aplicada2018prediccion}, Jordan \citep{ryanmosley2023algorithm}, India \cite{amnesty2022ban}, among others.

\end{itemize}

\begin{table*}[t]
\begin{center}
\begin{tabularx}{\textwidth}{XX}
 \textbf{Example Cases} (domain) &
\textbf{Description} \\ 
 \hline
  
 \hline
 \textbf{U.S. National Eating Disorder Assocation (NEDA)'s \emph{Tessa} Wellness Bot} \citep{wells2023eating} (chatbot) & Shortly after NEDA fired their hotline staff, activists posted on social media about how Tessa (NEDA's chatbot) gave harmful and triggering advice.\\
 
 \textbf{Facial Recognition in NYC Atlantic Plaza Towers} \citep{gagne2019how} (facial recognition) & Residents of a Brooklyn apartment complex partnered with a legal group to resist their landlord's plan to install facial recognition technology to ``secure'' access to the building.\\ 

 \textbf{The Spanish Ministry of the Interior's \emph{VioGén} Gender Violence Recidivism Risk Assessment} \citep{eticas2022audit} (family and social services) & The Spanish government's risk assessment algorithm under-estimated risks posed by perpetrators, withholding necessary protection services that led to adverse outcomes in domestic violence cases.\\

 \hline
\end{tabularx}
\caption{ \label{table:cases-maintext}Three example cases from our list of \caseno{} total case studies of real-world examples of campaigns to abandon an algorithm. See the complete list of \caseno{} cases in Appendix \ref{apdx:case-list}.}
\end{center}
\vspace{-4mm}
\end{table*}

\paragraph{Case set} From the cases that met the abandonment criteria, we selected a subset of \textbf{\caseno{}} as our case studies for our thematic analysis. 
We chose this subset based on several criteria, including (a) coverage of different domains, (b) access to enough sources to gain more information about a case to do a thematic analysis, and (c) reaching saturation in finding new themes in the analysis process. 
The selected cases included algorithms used in social media, chatbots, hiring, family and social services, benefits allocation, healthcare, education, property tech, and law enforcement.
We chose among cases within the same domain based on the amount of public information and documentation that was available for us to construct accurate event timelines.
We performed our thematic analysis in parallel to identifying relevant cases, and stopped adding new cases to the subset once we had reached saturation in finding new themes (\ie{} the types of events and factors relevant to abandonment).
Our final case set contains \caseno{} different algorithmic systems, \abandonedno{} of which had been abandoned at the time of writing\footnote{Some algorithms that we study were deployed in multiple different contexts (\eg{} PredPol was used by dozens of U.S. police departments \citep{haskins2019dozens}), and were abandoned in some, but not others. We count products like PredPol as a single ``algorithmic system'', and count them as being ``abandoned'' if they were abandoned in at least one such context.
}, while the remainder were still in use. 
\globalno{} out of the \caseno{} total cases involved campaigns that took place in countries outside of the U.S.
To provide context, we describe three example cases in Table \ref{table:cases-maintext} and a complete list in Appendix \ref{apdx:case-list}.
These \caseno{} cases are not intended to be a comprehensive list of all algorithms that were called to be abandoned, but rather serve as a starting point to conceptualize algorithm abandonment, illustrate its dynamics, and map out its implications.

\xhdr{Thematic analysis}
We gathered and reviewed public resources about each case, including news articles, research publications, social media posts, and any other relevant primary sources (\eg{} algorithm source code or court documents). 
For each case, we constructed a timeline by noting the month and year of important events. 
Our goal in noting important events was not to be comprehensive, but rather to provide ``good enough'' coverage of the types of events leading to abandonment, a standard used in past qualitative research \citep{yildirim2023creating}.
We reviewed sources to code metadata about the algorithm, such as its application domain, the type(s) of algorithmic harm \citep{shelby2023socio}, and other information relevant to whether it was ultimately abandoned (more details in Appendix \ref{apdx:methods-extended}).
We walk through the set of sources and resulting codes and event timeline for an example case in Appendix \ref{sec:apdx-analysis-extended}.

We conducted a bottom-up thematic analysis \citep{braun2006using} of case event timelines and metadata in a series of meetings among research team members.
In our thematic analysis of event timelines, we paid particular attention to the sequencing of key events, the time lapsed between them, and how each event contributed to the final decision of whether the system was abandoned.
Our analysis yielded six high-level types of events (or ``phases'') as the dynamics of the algorithm abandonment that we describe in Section \ref{sec:dynamics}.
To identify factors relevant to abandonment presented in Section \ref{sec:discussion}, we noted differences among algorithms that were abandoned almost immediately (\eg{} within days of their release), versus algorithms that took months or even years to be abandoned. 
We organized these differences into higher-level categories that resulted in the final seven factors.
We also noted different barriers that advocates of abandonment faced when interacting with algorithm owners (\ie{} asserting their role in this accountability relationship \citep{metcalf2023taking}).

\xhdr{Limitations}
In our study of the events that influenced abandonment, we analyzed \emph{publicly available} materials such as news articles or information obtained through public records requests.
We speculate, however, that there were other reasons why organizations chose to abandon the algorithm beyond those named in public sources, such as pressure from company shareholders \citep{whittaker2019amazon, klar2023meta} or financial projections \citep{puroncid2022are, mok2023ai}.
Future work that engages more directly with people involved in making the decision to abandon an algorithm may shed light on their motives.

\section{The Dynamics Toward Algorithm Abandonment}\label{sec:dynamics}
We explore the \emph{dynamics} of abandonment by proposing a process-oriented view of the events leading to abandonment. 
We visualized these dynamics for individual cases by creating a \emph{timeline of events} (shown in Figure \ref{fig:timelines}) that occurred until a system was abandoned (or until the time of writing for systems that were still in use).
In our thematic analysis, we found that most algorithms in our case set were abandoned \emph{after} they were deployed. However, our case set also includes cases of algorithms abandoned earlier in their development lifecycle. These cases provide meaningful comparison points to consider how algorithmic harms might be anticipated and prevented before a system is released.

\xhdr{The 6 D's of abandonment}  In our analysis of event timelines for the \caseno{} identified cases, the process leading to abandonment involved events that fell into six iterative phases of (1) \textbf{discovery}, which initiated critique and calls to abandon the system by the public, impacted communities, or their advocates, 
(2) \textbf{diagnosis}, where these actors gathered further evidence about the system and its impacts, 
(3) \textbf{dissemination}, where these actors amplified their concerns to a wider audience,
(4) \textbf{dialogue} between critics of the system and \owner{}s, until they reached a
(5) \textbf{decision} about whether to repair or abandon the system. 
We discuss how many abandoned systems still continued to impact peoples' lives in the final
(6) \textbf{death} phase, in which \owner{}s transitioned away from the algorithm and those impacted sought justice for past harms. 
Note that all six phases may occur in any stage of an algorithm's lifecycle (\eg{} design, development, usage), and are still relevant for algorithms that are not yet developed or deployed. In practice, the ``fall'' of an algorithm often proceeds \textit{iteratively} and \textit{non-linearly} throughout these six phases. 
Similarly, not all six phases may occur for every case. Regardless, this process-oriented view provides a useful lens to examine what types of events were influential across a range of cases.

\vspace{-2mm}
\subsection{Discovery}\label{sec:discovery}

Many cases had a catalyzing event where information that was previously unknown about the algorithm was made public and initiated wide-scale critique. We group these discoveries into two categories: (1) discovering the \emph{existence} of a previously obscured algorithmic system; and (2) discovering potential or realized algorithmic \emph{harms}.

\xhdr{Discovery of existence}  In some cases, criticism began once the algorithm's mere existence was made public. 
While \owner{}s could in theory provide public notice of their intentions to develop or use an algorithm in early stages of its lifecycle, many did not provide notice until the system was already in use. For example, the City of Pittsburgh ``\emph{experimented with}'' using a predictive policing algorithm for over a year before the public was made aware of its existence through the release of a public report \citep{capp2020primer}. 
Several other algorithms such as the New York Police Department's contract with facial recognition software vendor Clearview AI \citep{haskins2021nypd} evaded critique for months due to a lack of public notice.
Discoveries of an algorithms' existence often sparked criticism about a lack of community engagement and oversight \citep{capp2020primer} and the values they encoded (\eg{} through their task formulation or deployers' stated goals) \citep{hill2020secretive,logics2020safe,khan2021police}. 

\xhdr{Discovery of algorithmic harm} In other cases, initial critiques began after a discovery of potential or realized algorithmic harms. 
To understand what types of harms motivated calls to abandonment, we mapped all mentions of harms in case materials to the five major types of sociotechnical harms in \citet{shelby2023socio}: representational, allocative, quality-of-service, interpersonal, and social system harms.
We provide complete definitions and examples of each harm type Appendix Table \ref{tab:harm-defns}.
In summary, the algorithms in our case set spanned all five of \citet{shelby2023socio}'s harm categories. 
Algorithms deployed in similar domains resulted in similar harms, \eg{} chatbots could demean and stereotype \citep{kwon2021ai,kraft2016microsoft}, and resource allocation algorithms could result in opportunity loss \citep{ryanmosley2023algorithm,btah2016arkansas}. 
Several different types of harm often co-occurred for each single algorithmic system.\footnote{For example, San Diego's TACIDS algorithm \citep{tacids2021automated, williams2015facial} matched photographs of suspects against mugshot databases, where people with previous law enforcement encounters were more likely to be (falsely) flagged (representational harm). TACIDS was used to aid immigration enforcement efforts that separated families (interpersonal harm), and had implications for residents' constitutional rights (social system harms).}

Importantly, critics identified harms that implicated more than just the algorithm's \emph{functionality}, or ability to work as intended \citep{raji2022fallacy}. 
In other words, simply improving the algorithm to be more accurate, \ie{} to reduce what \citet{shelby2023socio} deem ``Quality-of-Service'' performance disparities, would not redress all harms. 
A more effective surveillance technology, for example, could still contribute to civil rights violations or interpersonal harms, such as accelerating incarceration or family separation at scale \citep{wang2022american, tacids2021automated}. 
While many cases we considered did involve dysfunctional algorithms, the remaining taxonomy categories remind us that other harms exist too.

\subsection{Diagnosis}\label{sec:diagnosis}
As the number of people aware of an algorithms' existence and potential harms grew, so did their capacity to \emph{diagnose} its potential to cause (further) harm.
We group critics' investigations into two categories: audits that analyzed the algorithms' \emph{outputs}, and investigations into \emph{human} decisions (such as design choices) surrounding the algorithm.

\xhdr{Examining the output: algorithm auditing}
Critics attempted to better understand the behavior of an algorithm by interacting with or probing it directly to observe its output. 
Many such audits involved querying the algorithm on several (real or simulated) inputs, and measuring properties of its output.
Large-scale targeted audits were a highly effective tool that precipitated decisions to abandon an algorithm.
For example, many organizations referenced the Gender Shades audit (which measured racial and gender accuracy disparities of commercial facial recognition algorithms \citep{buolamwini2018gender}) when they later instituted moratorium policies on the sale or usage of facial recognition technology \citep{buolamwini2023algorithmic}.

While large-scale audits were highly effective, in several cases, critics faced significant obstacles to conduct such audits. 
One dimension that significantly influenced the auditability of an algorithm was access to the model, as algorithms that were administered by a decision-maker internal to the organization (\eg{} a social worker assessing child welfare cases \citep{ho2020algorithm}) could not be queried at will. 
Independent auditors also often lacked access to realistic and representative data from the population of interest.
In some cases, the public was even unaware of what information the algorithm took as input to make decisions. 
For example, in Jordan, government officials refuse to release the ``socio-economic indicators'' of poverty that an algorithm which allocated cash assistance to citizens took into consideration \citep{hrw2023automated}.

To overcome these obstacles, in some cases critics worked together to collectively develop work-around strategies.
One strategy to obtain access to realistic input data was to design coordinated data collection efforts from consenting people who would be or were already impacted by an algorithmic system.
In Missouri, a coalition of legal aid organizations and third-party researchers partnered with home care nurses to fill out assessments to estimate a proposed home care allocation algorithm's impact for 1200 patients \citep{weil2023using}. 
In Spain, an independent nonprofit conducted phone interviews to understand the experiences of women who completed an algorithmic gender-based violence risk assessment that determined the police and legal resources allocated to them \citep{eticas2022audit}.
Successful data donation initiatives gave auditors access to population-level data that was critical to understanding the algorithm's behavior.

\xhdr{Examining the process: design decisions}
Algorithmic systems do not just ``come about'', but are rather the product of a series of \emph{decisions made by} designers, developers, and deployers \citep{selbst2024deconstructing, metcalf2023taking}. 
Critics attempted to understand algorithmic systems by examining the important decisions that brought them into being. 
For example, concerned Carnegie Mellon University students who opposed Pittsburgh's use of a predictive policing model filed a Right to Know request that revealed that the algorithm was rolled out without any meaningful consultation with the communities it was used to police \citep{capp2020primer}. 
Several advocacy groups have created resources to help people collect information about human decisions and processes surrounding algorithmic systems. 
For algorithms used by public agencies, organizations including the Electronic Frontier Foundation (``EFF'') \citep{eff2023foia} and Lucy Parsons Labs \citep{lpl2023who} have developed resources to help people navigate filing public records requests, which in some cases still only resulted in partial information being released \citep{capp2020primer, marx2021redacted}. 

\vspace{-2mm}
\subsection{Dissemination}\label{sec:dissemination}
Broader dissemination was a common strategy employed by critics to raise public awareness and put public pressure on \owner{}s. 
Dissemination sometimes caught the attention of elected officials or regulators who could apply pressure or exercise authority from above \citep{ho2020algorithm,warren2023yieldstar,ho2023child}. 
Some campaigns led by advocates of abandonment took place \emph{in-person} (offline), whereas in other cases campaigns took place \emph{online}. 
Other cases were supported by global media campaigns, such as the \#BanTheScan coalition supporting grassroots activist groups' efforts to ban government use of facial recognition technologies in NYC, Jerusalem, and Hyderabad \citep{amnesty2022ban}.
In many instances, only after the visibility of criticism increased did \owner{}s respond to and acknowledge the critique.

\xhdr{Online vs. Offline} While less common, a small number of cases involved in-person \emph{public demonstrations}.
All of the public demonstrations in our case set were organized to oppose systems that were intended to be deployed in a specific \emph{locality} (\eg{} in a particular city \citep{chang2018lapd}). 
Many such demonstrations were organized and attended by people who were directly impacted by the algorithm, such as student-organized protests against algorithmically assigned exam grades in the UK \citep{porter2020uk}, or a resident-led protest of a proposed facial recognition algorithm held on the sidewalk outside of their apartment \citep{elizalde2019brooklyn}. 
In contrast, criticism of tools with spatially diffuse user bases took place \emph{online}.
Several cases gained visibility through high engagement on online platforms, such as Twitter, Instagram, or YouTube by going ``viral'', such as users' interactions with NEDA's Tessa Bot \citep{wells2023eating}, or the Korean chatbot Lee-Luda \citep{kwon2021ai}.
We found that highly visible cases (as approximated through engagement metrics, such as high reaction or re-posts counts) were significantly more likely to receive a timely response from the \owner{}. 

\xhdr{Getting press} In several cases, media coverage of criticism played a critical role in the fall of an algorithm. 
As one example, Oregon's DHS decided just two weeks after an Associated Press story \citep{ho2020algorithm} to halt a pilot on child maltreatment risk prediction that had been in progress for four years. 
US Senator Ron Wyden ``\emph{reached out}'' to the department immediately after the story to ``\emph{ask questions about [the tool's] racial bias}'' \citep{ap2022oregon}.
Similar reporting efforts helped activist efforts gain momentum, and sometimes caught the attention of those in positions of power who applied pressure from above.

\vspace{-2mm}
\subsection{Dialogue}\label{sec:dialogue}
When criticism about an algorithm became more publicly visible, actors who held power over the algorithm were called to respond in a public \emph{dialogue} with their critics. These dialogues occurred in sites ranging from those that are more formal or compulsory for \owner{}s (\eg{} in court \citep{btah2016arkansas,btah2017idaho,van2020landmark,pierson_pierson_2023,mari_2022}), to those that were more informal (\eg{} on social media \citep{yee2021image,dickey_2020,thevergeTwitterTaught,wells2023eating,theguardianSouthKorean}).
Some \owner{}s took weeks, months, or even years to acknowledge their critics and respond \emph{at all}.
In this section, we examine how \owner{}s (dis)engaged with criticism about the algorithm. 
While some algorithms were primarily criticized by the public, other algorithms found support from other actors beyond their owners, and sparked a broader discourse about whether or not they should exist.

\xhdr{Denial vs. acceptance of algorithmic harm}  
When presented with evidence of algorithmic harm, some \owner{}s responded by \emph{discrediting} the evidence presented to them, either by creating counter-evidence \citep{eubanks2018response} or ad-hominem attempts to discredit the critic \citep{hao2020two, xiang2023eating}. 
For example, when an Instagram user shared an infographic that described a harmful and triggering interaction with NEDA's Tessa Wellness chatbot\footnote{https://www.instagram.com/p/Cs1jp1pPkOs/}, NEDA's VP of Communications accused her of lying in the post's comment section \citep{xiang2023eating}.
Interestingly, in many cases \citep{hao2020two, xiang2023eating}, initial denials were later retracted or withdrawn due to public scrutiny. 
After the NEDA VP's comment sparked outrage from other users, she deleted her comment and posted that she ``\emph{retracted her statement}'' \citep{xiang2023eating}. This pattern of denial followed by retraction is a broader trend in responses to attempts for algorithmic accountability \citep{raji2022actionable}.

In contrast, rather than resist, some \owner{}s immediately acknowledged the validity of the critiques. For example, when Twitter users began to investigate if its image cropping algorithm was racially biased \citep{yee2021image}, Twitter immediately issued a public statement that they were ``\emph{prioritizing work}'' to remedy bias in their algorithms \citep{agrawal2020transparency}. 
Cooperative responses that immediately acknowledged critiques allowed organizations to take actions (including abandonment) to redress harms more quickly.

\xhdr{The public vs. private dialogue} One dimension that influenced whether an algorithm was abandoned was whether \owner{}s responded publicly. 
Cases that were resolved interpersonally rather than in public were significantly less likely to result in critiques being resolved or the system being abandoned.
For example, Zoom users observed in their personal experiences \citep{shen2021everyday} that Zoom's virtual background feature tended to be more dysfunctional for users with darker skin, prompting some to tweet about it, but Zoom did not respond until one of these tweets\footnote{https://twitter.com/colinmadland/status/1307111816250748933} went viral. 
Zoom corporate replied to the tweet to invite its author to a call. 
Later that week, the author of the tweet posted that he had ``\emph{had a great convo with [Zoom's Chief DEI Officer] [...] I have no reason to doubt that they are serious about fixing problems with virtual backgrounds [...] the problem is deep and nuanced}''. 
Yet, such private resolutions still leave the broader public in the dark about what commitments the \owner{} has made. 

\xhdr{Opposing the abandonment/supporting the algorithm} 
As discourse surrounding an algorithm became increasingly visible, not all people believed that the algorithm should be abandoned.
For example, calls to ban police use of facial recognition technologies worldwide were met with concern from law enforcement officials.
In the U.S., police unions and conservative advocacy groups invested resources in lobbying against bans of facial recognition technologies \citep{gavin2022privacy, mosley2023how}.
Similarly, in Uganda, local police and members of the majority party lauded the facial recognition systems they purchased from Chinese vendor Huawei for ``\emph{boosting security}'' \citep{wilson2019uganda}.
Beyond law enforcement officials, even citizens support the use of facial recognition in some contexts: for example, 60\% of surveyed Indian citizens supported the use of facial recognition to identify protesters \citep{common2023status}.
These examples demonstrate how the debate to abandon an algorithm often implicates the larger surrounding political contexts and historic ideals (\eg{} surveillance, nationalism, and criminalization) that animate new technologies.

\vspace{-2mm}
\subsection{Decision}\label{sec:decision}
Those who held power over the algorithm could \emph{make decisions} to intervene on its design, development, or usage. We group these decisions about the algorithm into 4 categories based on their outcome: whether the algorithm was \emph{not abandoned}, in which the \owner{}s could (1) \emph{repair} the algorithm, or (2) make no changes; or the algorithm \emph{was abandoned} (3) by the \owner{}'s choice, or (4) by law or policy. 
In some cases, multiple decisions were made, \eg{} many \owner{}s first attempted to repair an algorithm before later deciding to abandon it.

\xhdr{Repair} \Owner{}s took several actions to repair or remediate potential harms, including fixing or updating dysfunctional technology \citep{deliban2018comment, afst2019methodology, ross2021epic} or creating better infrastructure or policies around its use \citep{dastin2020rite, sisitzky2021nypd, ross2021epic}. For example, when independent investigations revealed flaws in Epic's sepsis prediction model (which was deployed in several hundred U.S. hospitals) \citep{wong2021external}, the company completely ``overhauled'' its original algorithm and released a second version that used completely different input and target variables, and recommended that the model be trained on a hospital's \emph{own} data rather than being applied off-the-shelf \citep{ross2021epic}. 
However, repairs that improved the technology by making it more functional or less biased did not redress harms that resulted from the algorithms' existence.

\xhdr{Abandoned by the \owner{}'s choice} In our analysis, the majority of systems that were abandoned were abandoned by choice, \ie{} by a decision made by an organization that developed or used the algorithm. 
To name a few, Korean start-up ScatterLab disabled their Lee-Luda chatbot \citep{kwon2021ai},
Twitter decided to stop using an algorithm to crop users' images \citep{chowdhury2021sharing}, 
and leadership at Oregon's Department of Human Services decided to stop using a child welfare call screening algorithm \citep{ap2022oregon}.
Relatively few \owner{}s released public statements with explanations or named reasons why they decided to abandon the algorithm. 
Public statements that we found often referenced algorithmic harms as a primary reason why the system was abandoned \citep{lee2016learning, chowdhury2021sharing, kwon2021ai, hoover2023eating}.

\xhdr{Abandoned through policy} 
Algorithms overseen by uncooperative \owner{}s could still be abandoned through another governing body creating or enforcing applicable laws.
Several algorithms in our case set were deemed illegal through the interpretation of \emph{existing laws} (including laws written without an algorithm in mind) in a court of law.
For example, a Dutch court ruled that their government must stop using an algorithm that falsely predicted that thousands of low-income residents had committed benefits fraud, after finding that the algorithm violated citizens' right to privacy in the European Convention on Human Rights \citep{aw2020dutch}. 
U.S. courts ruled Medicaid benefits allocation algorithms in violation of existing laws (\ie{} due process law and the Administrative Procedure Act) in both Arkansas and Idaho \citep{calo2021automated, eppink2023testimony}. 
Other algorithms were banned by existing governing bodies founded to protect citizens' rights, such as the U.S. Federal Trade Commission issuing a five-year ban on Rite Aid's use of facial recognition because of the technology's risks to consumers \citep{ftc2023rite}.

However, all past cases of systems abandoned through executive or judicial enforcement took a \emph{significantly longer time} (\eg{} several years \citep{deliban2018comment, eppink2023testimony}) before the decision to abandon the algorithm was reached. 
These cases also required significant investment: \eg{} the legal team representing Idaho Medicaid beneficiaries spent over \$40,000 on expert consultations and 2,000 attorney hours to secure a settled agreement \citep{eppink2023testimony}.

Algorithmic systems were also abandoned through the \emph{creation of new policies} by a governing body (\eg{} a state legislature or city council) that had jurisdiction over the algorithm's design, development, or usage. 
New laws that target algorithmic systems have been rapidly growing in countries across the world \citep{iapp2023global}. 
For example, several U.S. state legislatures have proposed bans on algorithms developed for specific purposes, including facial recognition \citep{mosley2023how}, betting or casino algorithms \citep{hanchett2022rhode}, employee monitoring \citep{wilkinson2023growing}, or providing mental health counseling \citep{dm2023mh}.
All relevant policies that we identified for our cases targeted banning the \emph{usage} (rather than the \emph{development}) of technology. 

\vspace{-2mm}
\subsection{Death}\label{sec:death}
The \emph{death} phase explores what happens \emph{after} an algorithm is abandoned. Because algorithm development often entails creating infrastructure (such as data collection pipelines or portable products) \citep{ehsan2022the}, we discuss how abandoned algorithms' impacts can ``live on'' by influencing future processes \citep{ehsan2022the}, or in ``reincarnated'' products deployed elsewhere. 
Our understanding of the afterlife of abandoned algorithmic systems  builds upon \citet{ehsan2022the}'s conceptualization of the \emph{algorithmic imprint}, which describes how an algorithm's impact can extend beyond its deployment, by influencing infrastructural conditions, organizational practices, and societal norms that continue on in its ``afterlife''.

\xhdr{Overriding past decisions} Past decisions issued by an algorithm can continue to negatively impact people who received adverse outcomes. In the majority of cases of algorithms that were abandoned after their deployment, we found that the \owner{}s did \emph{not} re-consider or override past (harmful) decisions made using the algorithm. For example, to our knowledge, rejected resumes were not re-screened \citep{dastin2018amazon} and closed child welfare investigations were not re-opened \citep{ap2022oregon}. 
In a few notable cases, however, decisions were overturned and re-issued en masse. 
For example, in response to global protests against algorithmically-assigned A-level grades, U.K. exam boards globally rescinded all algorithm-assigned grades and revised them based on teacher-assessed grades (without the algorithmic standardization under criticism \citep{ehsan2022the}). 
Similarly, a legal settlement in Idaho guaranteed that people whose home care benefits were cut by an algorithm would continue to receive their original budgets \citep{btah2017idaho}. 

\xhdr{Making reparations} Beyond the rare re-issuing of decisions, some \owner{}s provided other kinds of reparations to symbolically or materially mend past harms. Material reparations, for example, may be especially appropriate for systems that caused allocative harm that made individuals incur additional financial costs, \eg{} by withholding access to healthcare benefits or employment. The majority of \owner{}s did not provide financial reparations for the harm caused, with a few notable exceptions. 
The Dutch government paid €30,000 to all families falsely accused of committing benefits fraud by an algorithm \citep{erdbrink2021government}.
The Arkansas DHS was required to pay \$460,000 in a court settlement to three plaintiffs whose benefits were severely cut, but many other hundreds of beneficiaries who were similarly impacted by the algorithm were not compensated \citep{ly2023arkansas}. 
No \owner{}s in our case set provided reparations for other types of non-material harms (\eg{} representational, psychological, or interpersonal harms).

\xhdr{Reincarnation} A growing number of investigations uncover how abandoned technologies can re-appear in different contexts \citep{kalluri2023surveillance, loewenstein2023the, scott2023lapd}. 
We observed that when a city cancelled a contract with a technology vendor, the company often remained and continued to sell their product elsewhere \citep{haskins2019dozens, haskins2021nypd}. 
Even as cities such as Palo Alto decided to drop their contracts with the predictive policing company PredPol, public records requests revealed that PredPol still had (many previously undisclosed) contracts with dozens of U.S. cities \citep{haskins2019dozens}. 
In response to bad press, some \owner{}s rebranded \citep{stop2021ghosts}: \eg{} one month after the LAPD stopped using PredPol, the company changed its name to Geolitica and its slogan from ``\emph{The Predictive Policing Company}'' to ``\emph{Data-Driven Community Policing}'' \citep{bhuiyan2021lapd}. 
In 2023, SoundThinking (the company that created ShotSpotter, another prominent law enforcement algorithm) acquired IP from Geolitica, and still maintains versions of its predictive policing algorithms \citep{mehrotra2023maker}. 
The case of PredPol exemplifies how built technologies are valued as infrastructure that remains and can be recycled even after they are abandoned.

\section{Factors relevant to algorithm abandonment}\label{sec:discussion}

Our analysis of historic cases across a range of domains reveals several patterns, including common phases that took place toward abandonment. However, beyond identifying commonalities, we also noticed key \emph{differences} relevant to the outcome of whether the system was ultimately abandoned, and how long it took to get to that stage.
In this section, we single out these \emph{differentiating socio-technical factors}, which span characteristics of the technical system and the social systems that the algorithm was embedded within. We highlight the implications of each factor for several stakeholders. For example, because abandonment is an important step that \owner{}s can take to redress (potential) algorithmic harms, we discuss ways that \owner{}s can design algorithmic systems that \emph{can be more easily abandoned} at different stages of their lifecycle. We also discuss implications of each factor for other external stakeholders, including FAccT researchers and oversight entities.

\xhdr{Users vs. Subjects} \emph{Is the algorithm a consumer product, or used on (unaware) subjects?} Algorithms in consumer products (\eg{} online platforms) "fell" more quickly than algorithms used by organizations (\eg{} governments) to analyze or make decisions about subjects who were potentially unaware of the algorithm’s existence. 
Direct users or consumers who interacted with technologies were more likely to be aware that there was an algorithm in place, and had more agency to resist harmful interactions with it. 
In cases where the direct user who consumed the algorithm's outputs used it to make decisions about someone else (\eg{} a policing agency using facial recognition on city residents), impacted individuals were often unaware of its use or existence, which prolonged the process leading to abandonment.
This distinction draws attention to the importance of understanding power differentials in who purchases and uses, versus who is most (negatively) impacted by algorithmic systems. 
\Owner{}s and critics alike should aim to empower and amplify the perspectives of those most impacted by algorithmic systems, even if they are not direct users.

\xhdr{Value proposition} \emph{What is the "value proposition" behind the use of the algorithm?} A critical factor relevant to cases' outcomes was understanding \owner{}s' motivation for creating or adopting the algorithm, \ie{} its value proposition. 
Publicly available value propositions often articulated a clear set of political values, \eg{} to reduce Medicaid spending (by cutting care) \citep{weil2023using}, to maximize returns to landlords (by raising rents) \citep{vogell2022rent}, or to enable constant and wide-scale surveillance \citep{dastin2020rite}.
In some cases, \owner{}s valued that automation allowed them to de-personalize decisions formerly made by humans, \eg{} that leasing agents or nurse assessors had ``\emph{too much empathy}'' \citep{vogell2022rent} to raise rents or cut hours.
When an algorithm's value proposition was clearly stated, in some cases critics organized successful campaigns that countered them (\eg{} challenging the notion that surveillance makes us feel safe \citep{logics2020safe}). 
\Owner{}s should make an algorithm's value proposition clear to the public. Critics should search for value propositions (\ie{} how is the technology being marketed to its buyers? \citep{roemmich2023values}) and make them visible so that they can be contested.

\xhdr{The "domino effect" caused by deep dependencies} \emph{How embedded is the algorithm into other systems? What other practices and processes "depend" on it?} Stand-alone algorithms released as independent products, like chatbots, were much easier to abandon than those embedded into larger social or technical systems, like algorithms used to automate important bureaucratic decisions \citep{alkhatib2019street}. 
Abandoning an algorithm that other components rely on might have a domino effect that requires changing its surrounding socio-technical components. 
For example, when courts ruled algorithms that assessed beneficiaries' eligibility illegal, public officials scrambled to develop a new process (\eg{} to train human assessors) to allocate care resources \citep{btah2016arkansas}. 
\Owner{}s should aim to minimize dependencies by creating and budgeting for back-up plans in a scenario where the algorithm is no longer in use, a proposal that has become popular in recent AI governance frameworks \citep{canadianaia, ostp2022blueprint}.

\xhdr{Procedural transparency and public notice} \emph{Is the public aware that an algorithm exists? Do we know how important decisions were made?} 
We found that many harmful algorithms were used for months (or even years) before they were discovered and the general public was made aware of their existence. 
In contrast, when \owner{}s made their plans to use an algorithm public \emph{before} it was deployed (like Missouri's NF LOC algorithm \citep{btah2017missouri}), in some cases they decided to abandon the algorithm before it could inflict harm in the real world. 
Thus, the extent to which \owner{}s made information about the algorithm public, and when, played a pivotal role in the time it took for harmful systems to be abandoned.
\Owner{}s should provide public notice and publicly release information about the algorithm's design at the earliest possible stages of the algorithm's lifecycle. 
FAccT researchers and advocacy groups can contribute to the small (yet growing) number of tools identified by \citet{inioluwa2024towards} designed to aid ``target identification'', \ie{} to ``\emph{uncover deployed systems by making them visible to external stakeholders}''. 
Finally, policymakers or organizations can institute transparency requirements for \owner{}s (\eg{} those proposed in \citep{reisman2018algorithmic,moss2021assembling,gehami2023when}).

\xhdr{Access and auditability}
\emph{What access do independent actors,} \eg{} \emph{members of the public, have to the algorithm?} As discussed in Section \ref{sec:diagnosis}, while independent audits were an extremely effective tool that triggered the decision to abandon an algorithm, independent actors had varying degrees of access to different algorithmic systems. FAccT researchers can contribute to infrastructure and build tools to support independent audits, such as those taxonomized by \citep{inioluwa2024towards}. One promising trend is that coalitions of different actors have started collaborations to share knowledge and data about their interactions with algorithmic systems, such as academic or non-profit research groups supporting public defenders and worker's unions \citep{weil2023using, btah2017missouri, fairfare}. Policymakers and advocacy groups can consider funding these independent audits \citep{mozilla2023auditing}.

\xhdr{Visibility of criticism and timely media coverage} \emph{How visible is the critique to other stakeholders?} As discussed in Sections \ref{sec:dissemination} and \ref{sec:dialogue}, critics' efforts to abandon an algorithm were often intertwined with a broader struggle for visibility.
Getting media coverage that uplifted critics' concerns and hopes for the future was often a turning point for accelerating an algorithm's fall. 
Other cases that went viral from organic engagement (such as amplification from influential users) were also significantly more likely to result in \owner{}s taking action to address algorithmic harms \citep{yee2021image, hoover2023eating}. Thus, journalists, media organizations, and ``everyday users'' \citep{shen2021everyday} can play an important role in advocating for abandonment.

\xhdr{Oversight and regulatory environment} \emph{Are policymakers receptive to creating new policies? Who is held liable for algorithmic harms? } As discussed in Section \ref{sec:decision}, instituting new ``bright-line rules'' \citep{ainow2023algorithmic} that explicitly ban specific types of algorithms takes far less time (and material investment) compared to interpreting existing laws in court. 
Policymakers can consider instituting such bans \emph{proactively}, even before the banned systems are in use, like the City of Somerville's ban on employees' use of facial recognition \citep{wu2019somerville}.
Another trend we observed is a lack of accountability under the law for vendors that develop algorithmic systems and sell them to organizations that oversee their deployment.
For example, when courts ruled Arkansas' roll-out of an algorithm developed by interRAI (an independent vendor) illegal, the state of Arkansas, not interRAI, was required to pay the full settlement amount to plaintiffs who were harmed \citep{ly2023arkansas}. 
This accountability structure means that even when one organization decides to stop using a harmful algorithm, it can continue to be used elsewhere with little consequence to the vendor who would continue to benefit from the algorithm \citep{lima2023who}. 
Policymakers should create and interpret existing laws to promote what \citet{metcalf2023what} calls ``\emph{distributed accountability}'', where vendors are also held liable for remedying algorithmic harms.

\section{Conclusion \& Future Work}

Grassroots movements, moratorium bans, and boardroom decisions to abandon harmful algorithmic systems continue to grow across the world.
Yet, comparatively little academic attention has been paid to understanding the phenomena of algorithm abandonment.
In our analysis of real-world cases involving (calls for) algorithm abandonment, we identified common phases of events that took place and key socio-technical factors that enabled (or prolonged) abandonment efforts.
Our analysis reveals power imbalances and information asymmetries that prevent critics from holding \owner{}s accountable.
Yet, victories from below in our case set illustrate how these power differentials might be challenged through policy interventions and collective action.
In our discussion of relevant factors, we highlight steps  that \owner{}s can take to create algorithms that can be more easily be contested and abandoned.
Regulators can consider making these steps mandatory, such as requiring \owner{}s to make important information public at early stages of design and development \citep{reisman2018algorithmic}.

Our preliminary work points to several exciting directions for future work on algorithm abandonment.
Researchers can take inspiration from successful collaborations that have built tooling and infrastructure to enable collective contestation, such as partnerships between academic researchers and worker's unions \citep{weil2023using, btah2017missouri, fairfare}.
Future research efforts can conduct more in-depth studies, such as interviews, to understand the experiences of actors involved in making the decision to abandon an algorithm, as well as other important stakeholders such as impacted communities, advocacy groups, and policymakers.
Future work can also center campaigns that go beyond calls to \emph{abandon}, and instead fight to \emph{abolish} harmful technologies.
In her conceptualization of ``\emph{abolitionist technologies}'', Ruha Benjamin writes that ``\emph{calls for abolition are never simply about bringing harmful systems to an end but also about envisioning new ones. After all, the etymology of abolition includes Latin root words for `destroy' (abolere) and `grow' (olere)}'' \citep{benjamin2019race}.
In our analysis, we found that algorithms in our case set implicate oppressive social systems or politics that informed their design.
How can the FAccT community can better support the many advocacy initiatives \citep{benjamin2019race,logics2020safe,milner2020abolish,stop2021ghosts} that work to contest the values that animate harmful algorithmic systems, and imagine more just futures?

\section{Impact Statements}\label{sec:impact-statements}
\xhdr{Researcher Positionality} 
We acknowledge that our lived experiences shape our positionality and perspectives \citep{liang2021reflexivity} in the study of algorithmic harms, which are often disproportionately faced by members of marginalized groups.
As researchers who were trained and work in the U.S., we recognize that our analysis, particularly our discussions of algorithmic accountability, contestation (\eg{} through public protests), and governance was influenced by the present political context of the U.S.
In our initial search for cases, we noticed that the majority of cases cited in FAccT research articles or reported in crowd-sourced AI incident databases occurred in the U.S.
Further, because our search was conducted in English, we likely missed relevant cases that occurred in non-English-speaking regions.
We attempted to address this regional bias by performing additional keyword searches and including an additional Latin America-based incident database \citep{varon2022not}.
We hope that our exploratory research raises academic awareness of algorithm abandonment as a \emph{global} phenomena, and inspires future work that more deeply interrogates similarities and differences relevant to how the dynamics of abandonment play out across regional contexts.

While all authors have experience studying algorithmic harm as academic researchers, it was important for us to better understand the needs and obstacles faced by activists taking action to resist oppressive technologies on the ground. 
Thus, as we conducted this research, we consulted experts who had lived experiences organizing to resist algorithmic harm.
Specifically, our research team met with members of the Surveillance Resistance Lab, a U.S.-based advocacy group founded to resist state and corporate surveillance technologies, who directed us to cases and offered feedback.
Looking forward, our research team plans to translate parts of this work to create resources that can directly benefit advocacy groups.

\xhdr{Ethical Considerations}  
In proposing a process-oriented view of the dynamics of algorithm abandonment, our analysis reveals commonalities that place seemingly dissimilar cases (of chatbots \citep{hoover2023eating}, B2B SAAS products \citep{vogell2022rent}, and carceral technologies \citep{tacids2021automated}) in conversation with one another.
These similarities enable us to formulate recommendations for future research and policy interventions related to algorithm abandonment.

However, our analysis has several limitations.
In focusing on what is \emph{shared} between a large and diverse set of algorithmic systems that are deployed in a wide variety of political contexts, we exclude important details that illustrate the uniqueness and complexity of each case.
Further, given the rapidly evolving nature of algorithmic capabilities and harms, we acknowledge that our process-oriented view may fail to describe the events leading to the fall of all algorithmic systems (including recent generative, multi-modal, and multi-purpose AI systems).
Because of these concerns, the 6 D's of abandonment are \emph{not} meant to serve as a model of steps for critics to follow, but rather provide descriptive context to understand similar historic struggles.

\xhdr{Adverse Impact} Our aim in identifying factors that permit (or hinder) algorithm abandonment is to support actors \emph{from below} in contesting harmful algorithmic systems.
To do so, we name and draw attention to common actions that actors \emph{from above} took to dodge and discredit their critics.
However, this work also has the potential to be misused by \owner{}s who model their actions after historic cases that successfully reinforced oppressive power structures (\eg{} by intentionally obscuring an algorithm).
Despite this risk, we advocate for the value of naming and identifying these tactics as a tool to strengthen the eco-system of accountability around algorithmic systems.

\begin{acks}

We thank Mizue Aizeki and Ed Vogel from the Surveillance Resistance Lab for offering feedback on this work.
We thank Ryan Steed, Meg Young, Upol Ehsan, and Sohini Upadhyay for directing us to related scholarship and cases.
We thank Alana Rogers and Ayush Tripathi for conducting preliminary research that informed this project.
HH and NJ acknowledge support from NSF (IIS2040929 and IIS2229881), PwC (through the Digital Transformation and Innovation Center), and the Block Center for Technology and Society at CMU. 
Any opinions, findings, conclusions, or recommendations expressed in this material are those of the authors and do not reflect the views of the National Science Foundation and other funding agencies.

\end{acks}




\appendix
\newpage
\section{Methodology (extended)}\label{apdx:methods-extended}

\paragraph{Keyword searches for international cases} To identify additional cases that met our criteria outside of the U.S., we conducted additional keyword searches on Google of names of regions/countries and our keywords from before, such as ``\emph{China algorithmic harm}''.

\subsection{Thematic analysis (continued)}\label{sec:apdx-analysis-extended}
\paragraph{Closed metadata coding} In addition to conducting a bottom-up thematic analysis where we performed \emph{open coding} to create event timelines, we also did some \emph{closed coding} where we recorded the below metadata for all cases:
\begin{itemize}
    \item Case domain
    \item Name of the organization overseeing the algorithm's deployment
    \item Name of any organizations involved with developing the algorithm (if different from deploying organization)
    \item Region(s) in which the algorithm is developed and deployed
    \item What does the algorithm take as input, and what does it predict?  How are these predictions then used to make decisions?
    \item Was the algorithm ever deployed?
    \item Who was (potentially) impacted by the algorithm?
    \item What were the algorithms' (potential) harms? (We later mapped these using the categories from \citet{shelby2023socio}'s taxonomy.)
    \item Who critiqued the algorithm, and why?
    \item Was the algorithm ever abandoned? For abandoned algorithms,
    \begin{itemize}
        \item How was the algorithm abandoned? (\ie{} by the \owner{}s choice? through policy?)
        \item Were there any caveats, \eg{} did the algorithm continue to be used in any other contexts?
        \item When in the algorithm's development lifecycle did abandonment occur?
        \item Is there a publicly available statement of why the algorithm was abandoned?
    \end{itemize}
\end{itemize}

We referred to answers to these questions explicitly when developing our process-oriented view (\eg{} in understanding the different harms that occurred, or mechanisms that algorithms were abandoned) and in identifying key socio-technical factors that differentiated algorithms that could be abandoned quickly.

\paragraph{Example annotation} For illustrative purposes, we walk-through how we coded an event timeline for an example case: the National Eating Disorder Association (NEDA)'s Tessa Wellness Bot (see the final timeline in Section \ref{apdx:timeline-descriptions} below). 
We became aware of this case by searching the AI Incident Database \citep{mcgregor2020preventing}. 
We began learning about the case by reading the news articles linked on its \href{https://incidentdatabase.ai/cite/545/}{AIID webpage}. 
While reading these articles, we closed-coded case metadata, and also noted any other important events that appeared to be relevant to the algorithmic harm(s) and decisions made by the model owners.
For example, we learned from a Vice article \citep{xiang2023eating} that shortly before the Tessa bot came under criticism for its potential to inflict harm, a news story broke that NEDA had decided to fire all of its hotline staffers following their decision to unionize, and intended to ``\emph{replace them}'' by instead directing people to Tessa, a chatbot that had been in operation since February 2022.
From this article, we also learned the name of the organization that developed Tessa, and gained insight into Tessa's design (\eg{} how users interacted with Tessa online). 
Two more recent articles linked on the AIID webpage \citep{wells2023eating, hoover2023eating} interviewed users who took to social media to criticize the prominent NEDA firing and the Tessa chatbot.
These articles directly linked to social media posts where we could directly engage with users' critiques and discussions about the algorithm, such as Instagram posts (\href{https://www.instagram.com/p/CtCa3_ZuMA0/}{link}).
We referenced these interviews and critiques to better understand the nature of the algorithm's potential for harm and who might be adversely impacted.
Finally, we also referred to these articles and primary sources (\eg{} NEDA's Instagram announcement, \href{https://www.instagram.com/p/Cs4BiC9AhDe/}{link}) to understand the organization's decision to abandon the algorithm. 
We conducted a final confirmatory Google search to check if the Tessa Bot (or any similar service created by Cass) is still in operation today.

\section{Event Timeline Figure (Expanded)}\label{apdx:timeline-descriptions}

We provide a detailed description of each labeled event in the event timeline visualization.\\

\textbf{Jordan's Takaful Cash Assistance Program} \citep{hrw2023automated, ryanmosley2023algorithm}
\begin{enumerate}
    \item May 2019, release. The Jordanian government launches the Takaful cash transfer program in partnership with funding from the World Bank.
    \item October 2021, discovery of harm. Human Rights Watch (an international NGO) begins conducting investigative interviews with Jordan residents about their experiences with Takaful (diagnosis).
    \item October 2022. Jordan's National Aid Fund (NAF), the social protection agency administering Takaful, participates in an interview with the Human Rights Watch (dialogue), in which they reveal that they changed the algorithm to stop predicting the benefit \emph{amount} after complaints that this was "arbitrary and confusing" (decision to repair). The algorithm now predicts benefit \emph{eligibility} instead.
    \item June 2023. The Human Rights Watch publishes the results of their investigation, which gets picked up and covered by several other large media outlets (dissemination). 

\end{enumerate}

\textbf{Metro21's Predictive Policing Pilot in Pittsburgh} \citep{capp2020primer, capp2020responding}
\begin{enumerate}
    \item February 2017, release. Pittsburgh begins a hotspot-based predictive policing pilot with researchers at CMU (through the Metro21 Smart Cities Institute). They do not notify or involve the public in the development of the program.
    \item October 2018, discovery of existence. CMU researchers publish a paper about an evaluation of their predictive policing pilot. The public is made aware of its existence.  A concerned group of CMU students begin to organize and file records requests (diagnosis).
    \item June 2020, In the wake of mass protests after the murder of George Floyd, CMU students form the Coalition Against Predictive Policing, where they hold virtual teach-ins and create a petition to the city (dissemination). The petition gathers 3.6k signatures. The City states that the "Metro21 pilot has ended" and that they are no longer using predictive policing.
    \item September 2020, Pittsburgh City Council votes to ban predictive policing in perpetuity (decision, death).
\end{enumerate}

\textbf{The National Eating Disorder Assocation (NEDA)'s Tessa Chatbot} \citep{wells2023eating, hoover2023eating, xiang2023eating}
\begin{enumerate}
\item February 2022, release. NEDA releases Tessa, a free AI "wellness bot" to help users "discover coping skills" for eating and body concerns.
\item May 2023, discovery of potential harm. NPR runs an exposé on how NEDA fired all helpline staff following their decision to unionize, and intends to "replace them" with Tessa.
\item May 2023. Five days after the NPR exposé, a fat-positive activist shares screenshots of harmful interactions with Tessa, in which the bot gave advice that might  trigger an eating disorder (\eg{} to count calories and lose weight). Her post goes viral (dissemination). A NEDA VP comments on the author's post that she is lying and asks the author to provide her with more evidence that the interaction actually occurred (dialogue).
\item May 2023. Two days later, NEDA announces that Tessa will be shut down (decision, death).

\end{enumerate}

\setcounter{table}{1}
\begin{table*}[h!]
\small
    \centering
    \begin{tabular}{l p{4.8cm} p{5.4cm}}
    \hline
    \textbf{Harm Category}             & \textbf{ Definition}                                                                                                    & \textbf{ Examples}                                         \\ 
    
    \hline
\textbf{Representational}       & Socially constructed beliefs and unjust hierarchies about social groups are reflected in system inputs and outputs.
 & Microsoft's Tay Chatbot \citep{thevergeTwitterTaught} began tweeting offensive content after interacting with trolls (\emph{demeaning, stereotyping}).\\ \hline

\textbf{Allocative}       & A system withholds information, opportunities, or resources from historically marginalized groups; often in domains that affect material wellbeing.
 & Arkansas' RUGS algorithm \citep{deliban2018comment} results in unexpected and severe cuts to beneficiaries' home care hours (\emph{opportunity loss}). \\ \hline

\textbf{Quality of Service}       & The system is dysfunctional, or underperforms disproportionately for certain groups of people along social categories of difference like disability, ethnicity, etc.
 & Zoom's Virtual Background feature \citep{dickey_2020} is dysfunctional for users with darker skin (\emph{alienation, increased labor}). \\ \hline

 \textbf{Interpersonal}       & The system adversely shapes relations between people or communities.
 & The Allegheny County Family Screening Tool (AFST) \citep{gerchick2023devil} assigns higher risk scores to parents with developmental disabilities, which may influence Allegheny DHS to open an investigation or deem them negligent, resulting in separation (\emph{loss of agency/control, privacy violation}). \\ \hline 

 \textbf{Social System}       & The system has adverse macro-level effects to knowledge systems, cultural impacts, political or civic systems, economic systems, or the environment.
 & San Diego's TACIDS algorithm \citep{williams2015facial} was used by cops or immigration officers to scan and instantly identify people, often without their consent (\emph{political and civic harms}). \\ 
\\
 \hline
\end{tabular}
\\
\caption{We map reports of potential or realized harms from the algorithmic systems we identified, to the five major types of sociotechnical harms from the taxonomy proposed by \citet{shelby2023socio}.  We also note the relevant harm "sub-types" for each case in italics, after a brief description of the harm that occurred. Definitions of each category (middle column) are taken verbatim from \citet{shelby2023socio}.}
\label{tab:harm-defns}
\end{table*}

\section{Case List}\label{apdx:case-list}

Interact with and download our case dataset online at this (\href{https://njohnson99.github.io/fall-of-algorithm-database/}{link})!

\begin{table*}[!ht]
    \centering
    \begin{tabular}{l p{3.7cm} p{8.7cm}}
    \textbf{}             & \textit{Case Name}                                                                                                    & \textit{Case Description}                                                                             \\ 
    \hline
\textbf{} & \textbf{Online platforms} & \\

\textbf{1} & Twitter Cropping Algorithm \citep{yee2021image} & Users hypothesized that an algorithm that cropped images to show feed previews compatible with device aspect ratios was more likely to crop out Black people (or other minority groups, like women) in images with multiple people.\\

\textbf{2} & Zoom Virtual Background \cite{dickey_2020} & Zoom's virtual background feature is more likely to be dysfunctional (\ie{} hide the body of or include parts of the background) for users with darker skin.\\   

 \hline
\textbf{} & \textbf{Chatbot} & \\

\textbf{3} &  Microsoft Tay Bot \cite{thevergeTwitterTaught} & Microsoft's Tay chatbot began tweeting offensive content after interacting with users who wanted it to do so.\\

\textbf{4} & U.S. National Eating Disorder Assocation
(NEDA)’s “Tessa” Wellness Bot \citep{wells2023eating} & Shortly NEDA fired all of their hotline staff (following their decision to unionize), activists posted on social media about how Tessa (NEDA's chatbot) gave harmful and triggering advice.\\ 

\textbf{5} & Meta's Galactica \cite{galacticaGalacticaDemo} & Within three days of releasing a public demo of their Galactica language model, Meta took down the public demo and re-branded Galactica as "available for researchers" only.\\ 

\textbf{6} & Scatter Lab's "Lee-Luda" Chatbot \cite{theguardianSouthKorean} & Seoul-based startup's AI-chatbot "Lee-Luda" was removed from Facebook messenger 20 days upon launch after users complained it was using offensive language, discriminatory and generated hate speech towards LGBTQ folks and people with disabilities. \\ 

 \hline
\textbf{} & \textbf{Employment} & \\
\textbf{7} & Amazon AI Recuriting Tool \cite{dastin2018} & Amazon disbanded an internal team whose goal was to train ML to screen resumes due to algorithmic gender bias. \\   

 \hline
\textbf{} & \textbf{Child Welfare} & \\

\textbf{8} & Oregon DHS's Safety at Screening Tool \citep{ap2022oregon} & Oregon DHS decided to stop using an AI-based screening tool (which is modeled after the AFST), in the wake of an April AP news article reporting that these tools have the ability to "heighten racial disparities in the child welfare system".\\

\textbf{9} & Pittsburgh Allegheny County Family Screening Tool (AFST) \citep{gerchick2023devil} & External researchers, journalists, and advocacy groups have raised concerns about the AFST, including that it is more likely to investigate (and ultimately deem negligent) (1) poor families, (2) Black families, and (3) people with disabilities. \\

\end{tabular}
\end{table*}

\newpage

\begin{table*}[!ht]
    \centering
    \begin{tabular}{l p{3.7cm} p{8.7cm}}

    \textbf{}             & \textit{Case Name}                                                                                                    & \textit{Case Description}                                                                             \\ 
    \hline
\textbf{} & \textbf{Child Welfare} & \\

\textbf{10} & Chile Local Childhood Office's Sistema Alerta Niñez (“Childhood Alert System”) \cite{valderrama2021sistema} &  Chile's Ministry of Social Development and Family  launched the program to strengthen childhood protection with an algorithm that provided risk scores used to determine necessary intervention, but rejected an independent external audit's results revealing the system wrongly analyzed variables associated with direct measures of poverty rather than safety, targeting and perpetuating over-intervention in poor neighborhoods. \\

\textbf{11} &  New Zealand's "Vulnerable Children PRM" (Predictive Risk Modeling) \cite{bukowitz_o’loughlin_2022} & New Zealand's Social Development Minister intervened and halted the deployment of the algorithm after seeing major concerns from additional research during the request for its ethics approval for further study. \\
\hline
\textbf{} & \textbf{Family \& Social Services} & \\

\textbf{12} & Teen Pregnancy Risk \newline {Prediction in Argentina} \cite{aplicada2018prediccion} & The Applied Artificial Intelligence Laboratory (LIAA) in the University of Buenos Aires published concerns in a detailed technical analysis and review of an algorithm developed through a partnership between Microsoft and province of Salta's Ministry of Early Childhood to predict teenage pregnancy. \\

\textbf{13} &  The Spanish Ministry of the Interior’s “VioGén”
Gender Violence Recidivism Risk  \newline{Assessment} \cite{eticas2022audit} & The Eticas Foundation partnered with victims to publish an external audit after the Spanish Ministry declined requests of auditing the algorithm that underestimated level of risk in domestic and intimate partner violence cases, leading to adverse outcomes. \\
\hline

\textbf{} & \textbf{Benefits} & \\

\textbf{14}  &   Arkansas Department of Human Services' Homecare Medicaid Program "RUGS" Assessment System \cite{btah2016arkansas} & Arkansas switched to using a new algorithm called "RUGS" to assess the hours of in-home care that residents are eligible for.  The algorithm resulted in severe cuts to the number of hours allocated to most residents. \\

\textbf{15} &   Missouri "NF LOC" Home Care \cite{btah2017missouri} & The State of Missouri released (in advance) their plan to switch to using an algorithm called "NF LOC" to assess eligibility for home-care services.  They released the algorithm for public comment, and external researchers found that it would result in severe cuts for a large group of residents (and also had other logical errors). \\

\textbf{16}  & Idaho Department of Health and Welfare's Care Recommendation Algorithm \cite{btah2017idaho} & The Idaho Department of Health and Welfare used an algorithm to assess individual Medicaid budgets available to people with disabilities in their home-based services waiver program. A class-action lawsuit, KW vs. Armstrong, resulted in the algorithm being made public, and the algorithm was deemed unconstitutional and "arbitrary". The case was settled so that people whose budgets were cut would continue to get their original budgets. \\

\end{tabular}
\end{table*}

\begin{table*}[!ht]
    \centering
    \begin{tabular}{l p{3.7cm} p{8.7cm}}

    \textbf{}             & \textit{Case Name}                                                                                                    & \textit{Case Description}                                                                             \\ 
    \hline
\textbf{} & \textbf{ Benefits} & \\

\textbf{17} &  World Bank's Automated Cash Transfer Program "Takaful" in Jordan \cite{toh2023automated} & Human Rights Watch published an investigation report that revealed how the World Bank's algorithm for distributing and ranking financial assistance through a cash transfer program spearheaded by Jordan government's social protection agency was wrongfully disqualifying those in need.  The algorithm used 57 factors that they declined to publicly disclose. \\

\textbf{18} & Netherlands' System Risk \newline{Indication} (SyRI) \cite{van2020landmark} & Local civil rights organizations filed a case against the state to stop using an algorithm to detect welfare benefits fraud. A Dutch court ruled that the algorithm violated the European Convention on Human Rights (ECHR) protections and halted its operation. \\

    \hline
\textbf{} & \textbf{ Healthcare} & \\

\textbf{19} & EPIC' Sepsis Prediction Tool \cite{ross2021epic} & Epic, an American healthcare software company (and prominent EHR vendor), launched the Epic Sepsis Model (ESM) in 2017, a proprietary linear model that alerts patients at high risk of sepsis. From 2021 onward, external and independent audits or investigations showed that the model was flawed; \eg{} it had lower AUC than reported, had a high false alarm rate, etc. \\

\textbf{20} & US HHS and UNOS (the United Network for Organ Sharing)'s "Acuity Circles Policy" \cite{washingtonpostLiverTransplant} & A joint investigation by the Washington Post and The Markup exposed how a new shift in the UNOS allocation algorithm's criteria in removing distance as a factor for organ donation and prioritizing sickest patients regardless of location, drastically decreased the number of transplant surgeries and increased deaths in poorer states. \\

\textbf{21} & UnitedHealth \& NaviHealth's "nH predict" Post-Acute Care Management \cite{pierson_pierson_2023} & Families of deceased beneficiaries filed a lawsuit in the U.S. District Court for the District of Minnesota against a insurance company's healthcare algorithm for determining post-acute care. The algorithm prematurely discharged patients from facilities against doctor recommendations and discontinued payment for necessary services and claims, targeting those on Medicaid. \\

\textbf{22} &  U.S. Health Systems' High-risk Healthcare Management" Prediction \cite{obermeyer2019dissecting} & Researchers at the UC Berkeley School of Public Health exploited a rich dataset that provided insight into a live, scaled algorithm that had been deployed nationwide, and found it perpetuated significant racial bias through using health care costs to determine who was most likely to benefit from care management programs.  \\

\textbf{23} &  New Zealand Health System's Equity Adjuster-Waitlist Tool \cite{perry_2023} & The ACT New Zealand political party and other clinicians started a petition against Auckland Health Minister's efforts to make healthcare more equitable for low-income, rural, and indigenous Maori and Pasifika communities through an algorithm that adjusted patients' position in elective surgery waitlists, claiming it was racial discrimination against other New Zealanders.  \\

\end{tabular}
\end{table*}

\begin{table*}[h]
    \centering
    \begin{tabular}{l p{3.7cm} p{8.7cm}}

    \textbf{}             & \textit{Case Name}                                                                                                    & \textit{Case Description}                                                                             \\ 

\hline

\textbf{} & \textbf{Property Tech} & \\

\textbf{24} & RealPage's YieldStar Algorithm \cite{vogell2022rent} & RealPage, a Texas-based company that provides property management software for landlords, developed an algorithm, YieldStar, that suggests daily prices for open rental units.  Critics claim that the software brings anti-trust concerns and has resulted in rent inflation across the U.S. \\

\hline

\textbf{} & \textbf{Education} & \\

\textbf{25} & UK Grading Tool \cite{meadway_2020} & In the 2020 pandemic, UK-based boards decided to use an algorithm to grade students' GCE A-level exams (international certificates which are important for university placements and scholarship opportunities). Worldwide protests broke out when the algorithmically calculated results were announced, which resulted in the grades being globally rescinded and re-assigned one week later. \\

\textbf{26}  & SAS Institute's Educational Value-Added Assessment \newline{System} (EVAAS) in Houston \cite{capo_bass_2017} & 
Houston Independent School District ruled in the settlement of a lawsuit brought by Houston and the American Federation of Teachers, to terminate the use of EVAAS, which used a student’s performance on prior standardized tests to predict academic growth in the current year, as basis for evaluating teacher effectiveness. \\

\hline

\textbf{} & \textbf{Facial Recognition} & \\

\textbf{27} & San Diego County Facial Recognition Tool (TACIDS) \cite{tacids2021automated} & San Diego County had a regional database known as the Tactical Idenfication System (TACIDS), which contained 1.8million booking photos and was used by 30 law enforcement agencies (including ICE).  Use of the database was ruled illegal by CA state AB 1215. \\

\textbf{28} &  NYPD using Clearview AI \cite{haskins2021nypd} & Released public records revealed that the NYPD had a formal relationship with facial recognition company Clearview AI, including a "trial" vendor agreement, despite the PD's remarks that they had never had a relationship with the vendor.  The NYPD released a formal facial recognition policy that banned use of Clearview in spring of 2020. \\

\textbf{29} & Detroit Project Green Light + DataWorks Plus \cite{hill_2020} & The Detroit PD signed a contract with a for-profit vendor (DataWorks Plus) to begin running facial recognition on live video streams from surveillance cameras (from the "Project Green Light" initiative), and other surveillance cameras (\eg{} drones). \\

\textbf{30} & City of Somerville's Face Surveillance Full Ban Ordinance \citep{wu2019somerville} & A unanimous (11-0) vote by Somerville City Council passed an ordinance to ban all use of FR technology by city employees (including, but not limited to, the city of Somerville Police). \\

\textbf{31} & Facial Recognition in NYC \newline{Atlantic} Plaza Towers \cite{gagne2019how} & Residents of a Brooklyn apartment complex partnered with a legal group to resist their landlord's plan to install facial recognition technology as a means to access the building. \\

\end{tabular}
\end{table*}

\begin{table*}[!ht]
    \centering
    \begin{tabular}{l p{3.7cm} p{8.7cm}}

    \textbf{}             & \textit{Case Name}                                                                                                    & \textit{Case Description}                                                                             \\ 
    \hline
\textbf{} & \textbf{Facial Recognition} & \\

\textbf{32} &  RiteAid's Enrollement Database for Surveillance Systems \cite{dastin2020rite} & RiteAid, a large U.S. drugstore chain, deployed facial recognition surveillance systems in over 200 stores.  Employees used the FR to surveil customers (\eg{} "persons of interest" who might attempt to engage in criminal activity). RiteAid stopped using FR after a Reuters investigation broke in 2020, and in 2023 were issued a five-year ban on their use of FR by the FTC. \\

\textbf{33} &  Hyderabad's \#BantheScan Campaign  \cite{amnesty2022ban} & India's Internet Freedom Foundation partnered with Amnesty International in kickstarting the \#BantheScan campaign globally and locally to resist law enforcement use of facial recognition. \\

\textbf{34} &  São Paulo Metro Intelligent Security System \cite{mari_2022}& The city judge of São Paulo ruled that the proposed facial reconigtion system did not meet the legal requirements under Brazil's General Data Protection Law (LGPD) in a lawsuit brought by the public defender and multiple human rights and civil advocacy organizations. \\

\textbf{35} &  Chinese Surveillance in the Xinjiang Uyghur Autonomous Region \cite{ohchr2022ohchr} & The Office of the United Nations High Commissioner for Human Rights published an assessment of concerns against police use of facial recognition technology to surveil and track Uyghur people across the Xinjiang region in China. \\

\textbf{36} &  New Zealand Identity Check Program  \cite{pennington_2023}& Data specialists and ethicists from New Zealand's indigenous Maori communities publicly commented and accused the government of ignoring their requests of involvement in improving the facial recognition technology with a 45\% failure rate for not only security surveillance, but also fraud detection. \\

\textbf{37} & Uganda's \newline{"Intelligent} Transport Monitoring System" \cite{nyeko_2023} & Human right organizations, local politicians, and other activists spoke out against the multi-purpose technology that will monitor traffic and public spaces by electronic license plate recognition, other high-density camera surveillance, and facial recognition for tracking vehicle locations for criminal activity and investigations, political opponents, and protests. \\

\hline

\textbf{} & \textbf{Predictive Policing} & \\

\textbf{38}  & UCLA PredPol \cite{lomibao_2021} & The LAPD began a 2011 contract with PredPol, Inc (which began as a research project with a UCLA professor), which they used for 9 years until the contract was not renewed in 2019.  A data leak showed that their algorithms disproportionately targeted Black, brown, and poor neighborhoods. \\

\textbf{39} & City of Pittsburgh \cite{capp2020primer} & Pittsburgh City Council voted to ban "obtaining, retaining, accessing, or using FR or predictive policing technology" after facing criticism from student activists for a predictive policing experiment in collaboration with researchers at Carnegie Mellon (Metro21). \\
\end{tabular}

\end{table*}

\begin{table*}[!ht]
    \centering
    \begin{tabular}{l p{3.7cm} p{8.7cm}}

\textbf{40} & Germany's Predictive Policing Software Program "HessenDATA" \cite{bundesverfassungsgericht_2023} & In a case brought by the brought by the German Society for Civil Rights (GFF), the German Federal Constitutional Court ruled that a predictive policing algorithm developed by U.S. company Palantir was unconstitutional on the basis of the right to informational self-determination under the country's fundamental privacy rights. \\

\hline
\end{tabular}

\end{table*}

\end{document}